\input harvmac.tex
\figno=0
\def\fig#1#2#3{
\par\begingroup\parindent=0pt\leftskip=1cm\rightskip=1cm\parindent=0pt
\baselineskip=11pt
\global\advance\figno by 1
\midinsert
\epsfxsize=#3
\centerline{\epsfbox{#2}}
\vskip 12pt
{\bf Fig. \the\figno:} #1\par
\endinsert\endgroup\par
}
\def\figlabel#1{\xdef#1{\the\figno}}
\def\encadremath#1{\vbox{\hrule\hbox{\vrule\kern8pt\vbox{\kern8pt
\hbox{$\displaystyle #1$}\kern8pt}
\kern8pt\vrule}\hrule}}

\overfullrule=0pt

%

\def\np#1#2#3{{\it Nucl. Phys.} {\bf B#1} (#2) #3}

\def\physrev#1#2#3{{\it Phys. Rev.} {\bf D#1} (#2) #3}

\font\zfont = cmss10 

\def\bigone{\hbox{1\kern -.23em {\rm l}}}
\def\ZZ{\hbox{\zfont Z\kern-.4emZ}}

\def\a{\alpha}
\def\b{\beta}
\def\g{\gamma}

\def\m{\mu}
\def\n{\nu}

\def\G{\Gamma}

\def\o{\over}

\Title{CALT-68-2186, hep-th/9807182}
{\vbox{
\hbox{\centerline{Complete Solution for M(atrix) Theory at Two Loops}}
\hbox{\centerline{ }}
}}
\smallskip
\centerline{Katrin Becker\footnote{$^\diamondsuit$}
{\tt beckerk@theory.caltech.edu} and Melanie 
Becker\footnote{$^\star$}
{\tt mbecker@theory.caltech.edu} }
\smallskip
\centerline{\it California Institute of Technology 452-48, 
Pasadena, CA 91125}
\bigskip
\baselineskip 18pt
\noindent

The complete result for the effective potential 
for two graviton exchange at two loops in 
M(atrix) theory can be expressed in terms of a 
generalized hypergeometric function.

\Date{July, 1998}

It has recently become clear that 
DLCQ 
supergravity and M(atrix) theory for finite $N$  
\ref\lfn{L.~Susskind, ``Another Conjecture about M(atrix) Theory'', 
hep-th/9704080.}
are in perfect agreement even at the level of three graviton exchange
\ref\oy{Y.~Okawa and T.~Yoneya,''Multibody Interactions of D Particles in 
Supergravity and Matrix Theory'', preprint UT-Komaba-98-13, hep-th/9806108.}.
Now more than ever before it will be important to have a 
systematic approach to calculations in 
M(atrix) theory, so that it is possible to go beyond computations
in general relativity.

In this note we will make one step in this direction. 
We will present the complete
result for the effective potential 
for two graviton exchange in M(atrix) theory at two loops.
Our notation and conventions are those of
\ref\bebe{K.~Becker and M.~Becker, ``A Two-Loop Test of M(atrix) 
Theory'', \np {506} {1997} {48}, hep-th/9705091, K.~Becker, M.~Becker, J.~Polchinski and A.~Tseytlin,
''Higher Order Graviton Scattering in M(atrix) Theory'', \physrev {56} {1997}
{3174}, hep-th/9706072, K.~Becker and M.~Becker, ``On Graviton Scattering Amplitudes
in M Theory'', \physrev {57} {1998} {6464}, hep-th/9712238. }.

The integrals that have to be solved for all diagrams appearing in
double graviton exchange 
at two loops in M(atrix) theory are (after the appropriate substitution)
of the type

$$ 
I=\int_1^{\infty} dx dy {(-1+x)^{\a} (-1+y)^{\b} \o (-1+x y)^{\g} }
x^{-\zeta+ \m}  y^{-\zeta + \n}. \eqno(1)
$$
Here $\zeta=b^2/2v$ and the other free parameters are constants whose
value depend on the concrete integral. 
This expression can be evaluated exactly and its solution is
$$ 
B\left(\a+1,\b+1\right) B\left(\zeta-\n-\b-1+\g,2+\a+\b-\g\right) 
$$ 
$$ 
_3 F _2 \left(1+ \a,2+\a+\b-\g , 1+\a+\m-\n;
2+\a+\b, \zeta +\a-\n+1 \right), \eqno(2)
$$
where $B(x,y)=\G(x) \G(y)/\G(x+y)$ is the beta function and $_3F_2$
is a generalized hypergeometric series of unit argument. 
Using the Stirling series
$$
\log \G (z)=
(z-1/2) \log z -z +1/2 \log( 2 \pi) +
\sum_{n=1}^{m} { B_{2n} \o 2n (2 n-1)} z^{2n-1}+
O(z^{-2m-1}), \eqno(3)
$$
as $z$ becomes large and the asymptotic expansion 
$$
_3F_2(a,b,c;d,e)=1+{abc \o de}+ \dots +
{(a)_n (b)_n (c)_n \o (d)_n (e)_n}+O(e^{-n-1}),\eqno(4)
$$
as $e$ becomes large (while $a,b,c$ and $d$ are fixed numbers), 
we can obtain the double expansion of the effective potential to 
all orders in the velocity and the impact parameter as $\zeta \to \infty$. 
Here $B_{2n}$ are the
Bernoulli numbers whose numerical values and recurrence relations 
can be found in Ramanujan \ref\rama{Ramanujan, ``Some Properties 
of Bernoulli's Numbers'', {\it Collected Papers} (1927),
Cambridge.}. 

The complete solution for the effective potential
coming from the fermions is:
$$
{\cal F}=
-{ 4 \sqrt{2} \pi \o v^{3/2}} 
{\G (\zeta +1) \o \G(\zeta +{3 \o 2})}
{_3 F_2} ({1\o 2}, {3 \o 2},{3 \o 2}; 3, \zeta +{3 \o 2})
-{ 135 \sqrt{2}  \pi \o 512 v^{3/2}} 
{\G (\zeta +1) \o \G(\zeta +{7 \o 2})}
{_3 F_2} ({5\o 2}, {5 \o 2},{7 \o 2}; 5, \zeta +{7 \o 2})+
$$
$$
{32 \sqrt{2} \pi \o v^{3/2}}
{(-\G({1\o 2}+\zeta)^2+\G(\zeta) \G(1+\zeta))\o
\G(\zeta) \G({1\o 2}+\zeta)}.\eqno(5)
$$
Incidentally, one of the hypergeometric series could be summed up
using Dougall's theorem.
The complete solution coming from bosons and ghosts is
given by the odd terms in $v$ of the formula
$$
{\cal B}=
{\pi \sqrt{2} \o 3 v^{3/2}} {\G(-{1 \o 2} +\zeta) \o \G( \zeta)}
\Bigl[ 
{49\o 8}{_3 F _2}({1 \o 2},{ 1\o 2},{1 \o 2};1 ,\zeta)-
{_3 F _2}({3 \o 2},{3 \o 2},{3\o 2}; 3, 1+\zeta)-
$$
$$
{137 \o 16} {(2 \zeta -1)\o \zeta} 
{_3 F _2}({ 1\o 2},{1 \o 2},{ 1\o 2}; 1, 1+\zeta)
\Bigr].\eqno(6)
$$
If we expand these formulas using the Stirling series we obtain the effective
potential computed in  {\bebe} 
up to order $v^8$. The leading order corresponds
to the relativistic correction of the $v^4$-term of
\ref\bfss{T.~Banks, W.~Fischler, S.~H.~Shenker and L.~Susskind,
``M Theory as a Matrix Model: A Conjecture'', \physrev {55} {1997} {5112},
hep-th/9610043.}, 
while the higher order terms correspond to
quantum gravity corrections.

The complete result 
of the effective potential of M(atrix) theory at one loop was found 
by DKPS
\ref\dkps{M.~R.~Douglas, D.~Kabat, P.~Pouliot and S.~H.~Shenker, 
``D-Branes and Short Distances in String Theory'',
\np {485} {1997} {85}, hep-th/9608024. }. 
It sounds plausible that the complete solution at 
one and two loops may teach us how to solve the model
for an arbitrary number of loops.

\vskip 1cm
\noindent {\bf Acknowledgement}

\noindent This work was supported by the U.S. Department of Energy 
under grant DE-FG03-92-ER40701.

\listrefs

\end